\documentclass[11pt,a4paper]{article}
\usepackage[hyperref]{acl2017}
\usepackage{times}
\usepackage{latexsym}

\usepackage{rotating}
\usepackage{amsmath}
\usepackage{booktabs}
\usepackage{array}
\usepackage{multirow}
\usepackage{graphicx}
\usepackage{color}
\usepackage{blindtext}
\usepackage{subcaption}
\usepackage[skip=4pt]{caption}
\usepackage{enumitem}
\usepackage{multirow}
\usepackage{tikz}
\usetikzlibrary{shapes,snakes,positioning}
\aclfinalcopy %
\def\aclpaperid{417}

\newif{\ifhidecomments}
\hidecommentsfalse

\ifhidecomments
    \newcommand{\chenhao}[1]{}
    \newcommand{\nascomment}[1]{}
    \newcommand{\dbccomment}[1]{}
\else
    \newcommand{\chenhao}[1]{\textcolor{blue}{[#1]}}
    \newcommand{\nascomment}[1]{\textcolor{red}{[#1]}}
    \newcommand{\dbccomment}[1]{\textcolor{cyan}{[#1]}}
\fi

\newcommand{\addFigure}[2]{\includegraphics[width=#1]{plots/#2}}
\newcommand{\para}[1]{\noindent{\bf #1}}
\newcommand{\figref}[1]{Fig.~\ref{#1}}
\newcommand{\equationref}[1]{Eq.~\ref{#1}}
\newcommand{\secref}[1]{\S \ref{#1}}
\newcommand{\tableref}[1]{Table \ref{#1}}
\newcommand{\ideas}{ideas\xspace}

\newcommand{\idea}{idea\xspace}

\newcommand{\USA}{U.S.\xspace}
\newcommand{\cooccurrence}{cooccurrence\xspace}

\newcommand{\cooccur}{cooccur\xspace}
\newcommand{\cooccurs}{cooccurs\xspace}
\newcommand{\Cooccurrence}{Cooccurrence\xspace}

\newcommand{\correlation}{prevalence correlation\xspace}
\newcommand{\Correlation}{Prevalence correlation\xspace}
\newcommand{\frequency}{prevalence\xspace}
\newcommand{\pmi}{$\widehat{\operatorname{PMI}}$\xspace}
\newcommand{\freqcorr}{$\hat{r}$\xspace}
\newcommand{\kwords}{keywords\xspace}

\newcommand{\Kwords}{Keywords\xspace}
\newcommand{\friends}{friendship\xspace}
\newcommand{\romance}{tryst\xspace}
\newcommand{\armsrace}{arms-race\xspace}
\newcommand{\headtohead}{head-to-head\xspace}
\newcommand{\Friends}{Friendship\xspace}
\newcommand{\Romance}{Tryst\xspace}
\newcommand{\Armsrace}{Arms-race\xspace}
\newcommand{\Headtohead}{Head-to-head\xspace}

\newcommand{\ideaname}[1]{{\small\sf #1}\xspace}
\newcommand{\ideapair}[2]{(\ideaname{#1}, \ideaname{#2})}
\definecolor{friendscolor}{RGB}{19, 126, 109}
\definecolor{armsracecolor}{RGB}{68, 142, 228}
\definecolor{headtoheadcolor}{RGB}{152, 0, 2}
\definecolor{romancecolor}{RGB}{207, 98, 117}
\newcommand{\formalideas}{\mathcal{I}}

\title{
Friendships, Rivalries, and Trysts:\\
Characterizing Relations between Ideas in Texts
}

\author{Chenhao Tan$^\ast$ \quad Dallas Card$^\dagger$ \quad Noah A.~Smith$^\ast$ \\
\begin{tabular}{cc}
$^\ast$Paul G. Allen School of Computer Science \& Engineering &
                                                                 $^\dagger$School of Computer Science \\
University of Washington  & Carnegie Mellon University \\
Seattle, WA 98195, USA &
Pittsburgh, PA 15213, USA
\end{tabular}
\\ {\tt \href{mailto:chenhao@chenhaot.com}{chenhao@chenhaot.com}}  \quad
{\tt \href{mailto:dcard@cmu.edu}{dcard@cmu.edu}}  \quad
{\tt \href{mailto:nasmith@cs.washington.edu}{nasmith@cs.washington.edu}}
}
\begin{document}
	
    \maketitle

\begin{abstract}
Understanding how \ideas relate to each other is a fundamental question
in many domains, ranging from 
intellectual history to public communication.
Because \ideas are naturally embedded in texts, we propose 
the first
framework
to systematically characterize the relations between ideas based on their occurrence in a corpus of documents, independent of how these \ideas are represented.
%
%
%
%
%
%
%
%
%
%
%
%
%
%
Combining two statistics---{\em cooccurrence within documents} and
{\em prevalence correlation over time}---our 
approach
reveals a number of different ways in which
 ideas can cooperate and compete.
For instance, two \ideas can closely track each other's \frequency over time, 
and yet rarely \cooccur,
almost like a ``cold war'' scenario.
We observe that pairwise \cooccurrence and \correlation
exhibit different distributions.
We 
further 
demonstrate that our approach is able to uncover intriguing relations between \ideas
through in-depth case studies on news articles and research
papers.

\end{abstract}

\section{Introduction}
\label{sec:intro}
%
%
%

%
\begin{figure*}[t]
    \centering

\begin{tabular}{c@{\hspace{0.02\textwidth}}c@{\hspace{0.02\textwidth}}|@{\hspace{0.02\textwidth}}c}
& {\large anti-correlated} & {\large correlated} \\
\parbox[t]{2mm}{\multirow{2}{*}{\rotatebox[origin=c]{90}{\large likely to \cooccur}}} & {\large \textcolor{romancecolor}{\romance}} & {\large \textcolor{friendscolor}{\friends}} \\
& \addFigure{0.3\textwidth}{data_exp/immigration/{immigration_lda_noaxis_romance_immigration,imm_immigration,det}.pdf}
& \addFigure{0.3\textwidth}{data_exp/immigration/{immigration_lda_noaxis_friends_immigrant,undoc_obama,president}.pdf}
\\
\midrule
\parbox[t]{2mm}{\multirow{2}{*}{\rotatebox[origin=c]{90}{\large unlikely to \cooccur}}} & {\large \textcolor{headtoheadcolor}{\headtohead}} & {\large \textcolor{armsracecolor}{\armsrace}}\\
& \addFigure{0.3\textwidth}{data_exp/immigration/{immigration_lda_noaxis_head-to-head_immigrant,undoc_illegal,alien,i}.pdf}
& \addFigure{0.3\textwidth}{data_exp/immigration/{immigration_lda_noaxis_arms-race_immigration,imm_republican,part}.pdf}\\
\end{tabular}
    \caption{
    Relations between ideas 
    in the space of \cooccurrence and \correlation (\correlation is shown explicitly and \cooccurrence is encoded in row captions).
    We use topics from LDA \citep{Blei:2003:LDA:944919.944937} to represent ideas.
    Each topic is 
    named
    with a pair of words that are most strongly associated with the topic in LDA.
    Subplots show examples of
    relations between topics found in U.S. newspaper articles on immigration from 1980 to 2016, color coded to match the description in text. 
    The $y$-axis represents the proportion of news articles in a year
    (in our corpus) that contain the corresponding topic.
    {All examples are among the top 3 strongest relations in each type except \ideapair{``immigrant, undocumented''}{``illegal, alien''}, which corresponds to the two competing narratives.}
    We explain the formal definition of strength in
    \secref{sec:framework}.
    \label{fig:intro_immigration}}
\end{figure*}

Ideas  exist in the mind, but are made manifest in language,
where they  compete with each other for the scarce resource of human attention.
\citet{milton1973areopagitica} 
used the 
``marketplace of ideas'' 
metaphor to 
argue that the truth will win out when ideas freely compete; 
\citet{dawkins2016selfish} similarly likened the 
evolution of \ideas 
 to natural selection of genes. 
We propose
a framework to quantitatively characterize
competition and 
cooperation
between \ideas in texts, independent of how 
they
might be represented.

By ``ideas'', we mean any discrete conceptual units that can be identified as being 
present or absent in a document.
In this work, we 
consider representing ideas using keywords and topics obtained in an unsupervised fashion, but our way of characterizing the \textit{relations} between ideas could be applied to many other types of textual representations, such as frames \citep{card2015media} and hashtags.
What does it mean for two \ideas to compete in texts, quantitatively? 
%
%
Consider, for example, the issue of immigration.
There are two strongly competing narratives about 
the roughly 11 million people\footnote{As of 2014, according to the most recent numbers from the Center for Migration Studies \cite{warren.2016.cms}.} who are residing in the 
United States 
without permission.
One is
``illegal aliens'', 
who ``steal'' jobs and deny opportunities to legal immigrants;
the other is 
``undocumented immigrants'',
who are already part of the fabric of society and deserve a path to citizenship \cite{merolla2013illegal}. 

Although prior knowledge 
suggests that these two
narratives
compete, it is
not immediately obvious 
 what measures might reveal this competition  in a corpus of writing
 about immigration.
One question is whether or not these two ideas \cooccur in the same documents. 
In the example above,
these narratives are used by distinct groups of people with different ideologies.
The fact that they don't \cooccur is one clue that they may be in competition with each other.

However, \cooccurrence is insufficient to express the selection process of \ideas, i.e., some \ideas fade out over time, while others rise in popularity, analogous to the populations of species in nature. 
Of
the 
two narratives on immigration, we may expect one to win out at the expense of another as public opinion shifts.
Alternatively, we might expect to see these narratives reinforcing each other,
as both sides intensify their messaging in response to growing opposition,
much like
the U.S.S.R. and the U.S. during the cold war. 
To capture these possibilities, we use \correlation over time.
Building on these insights, we propose a framework that combines \cooccurrence within documents and \correlation over time.
This framework gives rise to 
four possible types of relation that correspond to the four quadrants in \figref{fig:intro_immigration}.
We explain each type using examples from news articles in U.S. newspapers on immigration from 1980 to 2016. %
Here, we have used LDA to identify ideas in the form of topics, and we denote each idea with a pair of words most strongly associated with the corresponding topic.

\begin{itemize}[itemsep=0pt,topsep=1pt,leftmargin=0pt,label={}]

    \item {\bf \textcolor{friendscolor}{\Friends}} (correlated over time, likely to \cooccur). 
   The \ideaname{``immigrant, undocumented''} topic tends to \cooccur with \ideaname{``obama, president''} and both topics have been rising during the period of our dataset, likely because the
    ``undocumented immigrants'' narrative was an important part of Obama's 
    framing of the immigration issue
     \cite{haynes2016}.

    \item {\bf \textcolor{headtoheadcolor}{\Headtohead}} (anti-correlated over time, unlikely to \cooccur).
    \ideaname{``immigrant, undocumented''} and \ideaname{``illegal, alien''} are in a \headtohead competition: 
    these two topics rarely \cooccur, and \ideaname{``immigrant, undocumented''} has been growing in prevalence, while the usage of \ideaname{``illegal, alien''} in newspapers has been declining.
    This observation agrees with a report from Pew Research Center \cite{guskin2013illegal}.

    \item {\bf \textcolor{romancecolor}{\Romance}} (anti-correlated over time, likely to \cooccur).
    The two off-diagonal examples use 
    topics
    related to law enforcement.
    Overall, \ideaname{``immigration, deportation''} and \ideaname{``detainee, detention''} often \cooccur but \ideaname{``detainee, detention''} has been declining, while \ideaname{``immigration, deportation''} has been rising.
    This possibly 
   relates to the promises to overhaul the immigration
    detention system
    \cite{kalhan2010rethinking}.\footnote{The tryst relation is the least intuitive, yet is nevertheless observed. The pattern of being anti-correlated yet likely to cooccur is typically found when two ideas exhibit a friendship pattern (cooccurring and correlated), but only briefly, and then diverge.}

    \item {\bf \textcolor{armsracecolor}{\Armsrace}} (correlated over time,
      unlikely to \cooccur). 
    One of the above law enforcement topics (\ideaname{``immigration, deportation''}) and a topic on the Republican party (\ideaname{``republican, party''}) hold an \armsrace relation:
    they are both growing in \frequency over time, but rarely \cooccur,
    perhaps suggesting an underlying common cause. 

\end{itemize}

{\em Note that our terminology describes the relations between ideas \underline{in texts}, not necessarily between the entities to which the ideas refer.}
For example, we find that the relation between ``Israel'' and ``Palestine''
is ``\friends''  in news articles on terrorism, based on their \correlation and
\cooccurrence in that corpus.

We introduce the formal definition of our framework in \secref{sec:framework} and apply it to news articles on five issues and research papers from ACL Anthology and NIPS as testbeds. 
We operationalize \ideas using topics \cite{Blei:2003:LDA:944919.944937} and \kwords \cite{Monroe21092008}.

To explore whether the four relation types exist and how strong these relations are,
we first examine the marginal and joint distributions of \cooccurrence and \correlation (\secref{sec:observation}).
We find that \cooccurrence shows a unimodal normal-shaped distribution but \correlation demonstrates more diverse
distributions across corpora.
As we would expect, there are, in general, 
more and stronger \friends and \headtohead relations 
than \armsrace and \romance relations.

Second, we demonstrate the effectiveness of our framework through in-depth case studies (\secref{sec:explore}).
We not only validate existing knowledge
about some news issues and research areas, but also 
identify
hypotheses that require further investigation.
For example, using keywords to represent ideas, a top pair with the \romance relation in news articles on terrorism is ``\ideaname{arab}'' and ``\ideaname{islam}'';
they are likely to \cooccur, but ``\ideaname{islam}'' is rising
in relative \frequency while ``\ideaname{arab}'' is declining.
This suggests a conjecture that the news media have increasingly
linked terrorism to a religious group rather than an ethnic group.
We also show relations between topics in ACL that center around \ideaname{machine translation}.

Our work is a first step towards understanding relations between \ideas from text corpora, a complex and important research question.
We provide some concluding thoughts in \secref{sec:conclusion}.

\section{Computational Framework}
\label{sec:framework}

The aim of our computational framework is to explore \emph{relations} between ideas. %
We thus assume that the set of relevant \ideas has been identified,
and those expressed in each document have been 
tabulated.
Our open-source implementation is at \url{https://github.com/Noahs-ARK/idea_relations/}.
In the following, we introduce our formal definitions and datasets.

\subsection{Cooccurrence and Prevalence Correlation}

As discussed in the introduction, we focus on two 
dimensions to quantify relations between \ideas: 
\begin{enumerate}
\item \cooccurrence reveals to what extent two ideas tend to occur in the same contexts;
\item similarity 
between the relative \frequency of \ideas 
over time 
reveals
how two \ideas relate in 
terms of popularity or coverage.
\end{enumerate}
Our input is
a collection of documents, each represented by a set of \ideas and indexed by time.
We denote a {\em static} set of ideas as $\formalideas$ and a text
corpus 
that 
consists of these ideas as $C = \{ D_1, \ldots,
D_T\}$, where $D_t = \{d_{t_1}, \ldots, d_{t_{N_t}}\}$ gives the collection of documents at timestep $t$,
and each document, $d_{t_k}$, is 
represented as
a subset of \ideas in $\formalideas$.
Here $T$ is the total number of timesteps, and $N_t$ is the number of
documents at timestep $t$. %
It follows that the total number of documents $N = \sum_{t=1}^T N_t$.

\begin{figure}[t]
{
  \setlength{\abovedisplayskip}{0pt}
  \setlength{\belowdisplayskip}{\abovedisplayskip}
  \setlength{\abovedisplayshortskip}{0pt}
  \setlength{\belowdisplayshortskip}{0pt}
\begin{align}
\label{eq:pmi}
& \forall x, y \in \formalideas,
\widehat{\operatorname{PMI}}(x, y) =
    \log \frac{\hat P(x,y)}{\hat P(x) \hat P(y)} \nonumber\\
   & = C+ \log \textstyle \frac{ 1 + \sum_t\sum_k \boldsymbol{1}\{x \in d_{t_k}\}
  \cdot \boldsymbol{1}\{ y \in d_{t_k}\}  %
  }{
  \left(1 + \sum_t\sum_k
  \boldsymbol{1}\{x \in d_{t_k}\}\right) \cdot  \left(1 + \sum_t\sum_k
  \boldsymbol{1}\{y \in d_{t_k}\}\right) } \\ 
\label{eq:pearson}
& \hat{r}(x, y) = \textstyle \frac{\sum_t \left( \hat{P}(x \mid t) - \overline{\hat
    {P}(x \mid t)} \right) \left( \hat P(y \mid t) - \overline{\hat
    {P}(y \mid t)} \right) }{ \sqrt{ \sum_t \left( \hat P(x \mid t) - \overline{\hat
    {P}(x \mid t)} \right)^2 } \sqrt{ \sum_t \left( \hat P(y \mid t) - \overline{\hat
    {P}(y \mid t)} \right)^2} }
\end{align}
}%
\caption{Eq.~\ref{eq:pmi} is the empirical pointwise mutual information for two ideas, our measure of
  \cooccurrence of \ideas;  note that we use add-one smoothing in
  estimating PMI.
   Eq.~\ref{eq:pearson} is the Pearson correlation
  between two ideas' 
  \frequency
  over time.
  }
  \label{fig:eqs}
  %
  %

\end{figure}

In order to formally capture the two dimensions above, we employ two
commonly-used statistics.
%
%
%
First, we use empirical pointwise mutual information (PMI) to capture the \cooccurrence of \ideas within the same document 
\cite{Church:1990:WAN:89086.89095}; see \equationref{eq:pmi} in  \figref{fig:eqs}.
Positive $\widehat{\mathrm{PMI}}$ indicates that \ideas occur together more frequently than 
would be 
expected
 if they were independent,
 while negative $\widehat{\mathrm{PMI}}$ indicates
 the opposite.

Second, we compute the correlation between normalized document frequency of \ideas to
capture the relation between the relative \frequency of \ideas across
documents over time; see \equationref{eq:pearson} in \figref{fig:eqs}.
Positive \freqcorr indicates that two \ideas have similar
\frequency over time,  while negative \freqcorr suggests 
two anti-correlated
\ideas 
(i.e., when one goes up, the other goes
down).

The four types of relations in the introduction can now be obtained
using \pmi and \freqcorr, which capture \cooccurrence and \correlation
respectively.  
We further define the \textit{strength} of the relation between two ideas as
the absolute value of the product of their \pmi and \freqcorr scores:
\begin{equation}
\forall x, y \in \formalideas,  \operatorname{strength}(x, y) =
|\widehat{\operatorname{PMI}}(x, y)  \times \hat{r}(x, y)|.
\label{eq:strength}
\end{equation}

\begin{figure*}[t]
    \centering
    \begin{subfigure}[t]{0.32\textwidth}
        \addFigure{\textwidth}{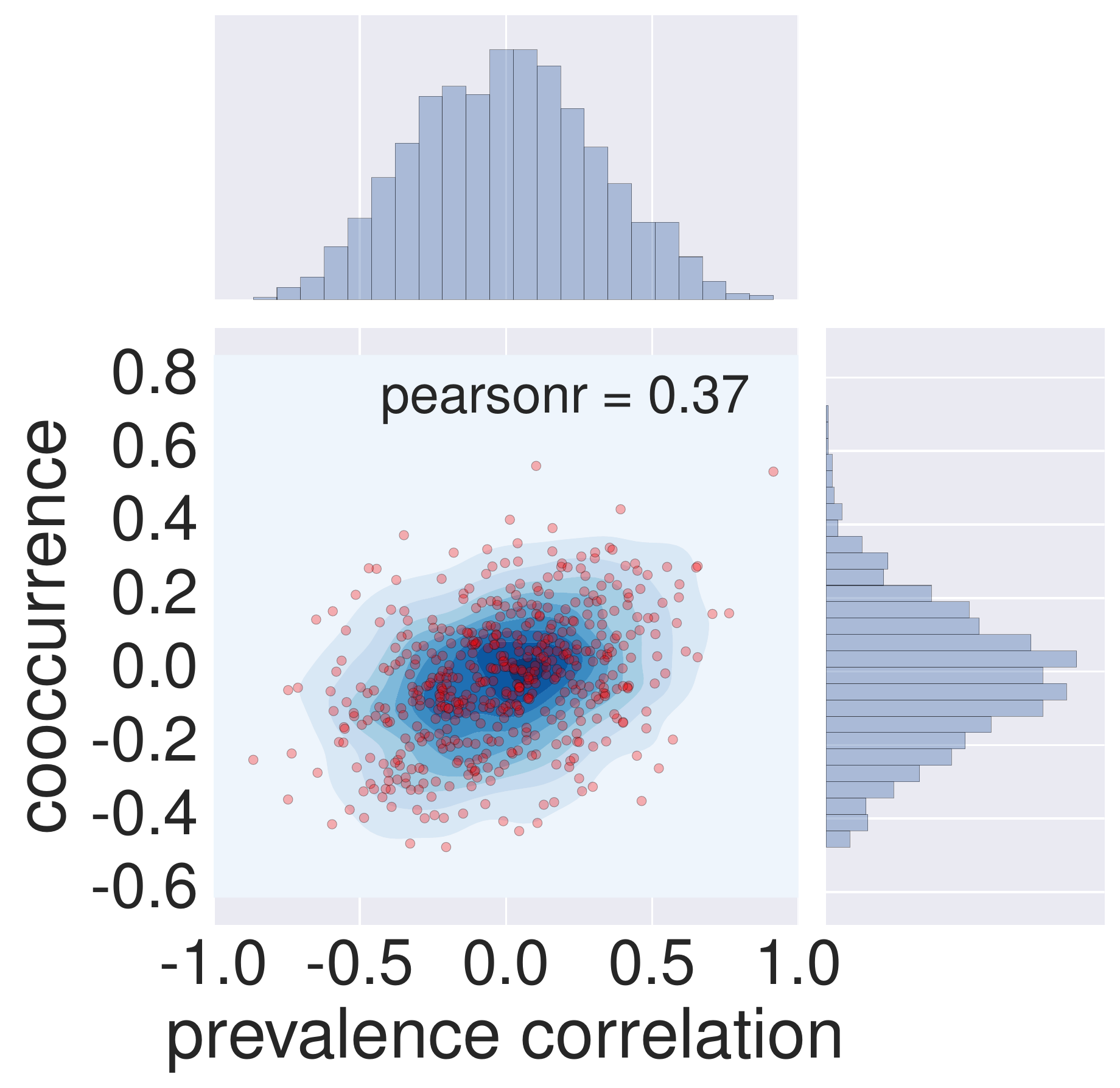}
        \caption{Terrorism topics}
        \label{fig:dist_terrorism_lda}
    \end{subfigure}
    \hfill
    \begin{subfigure}[t]{0.32\textwidth}
        \addFigure{\textwidth}{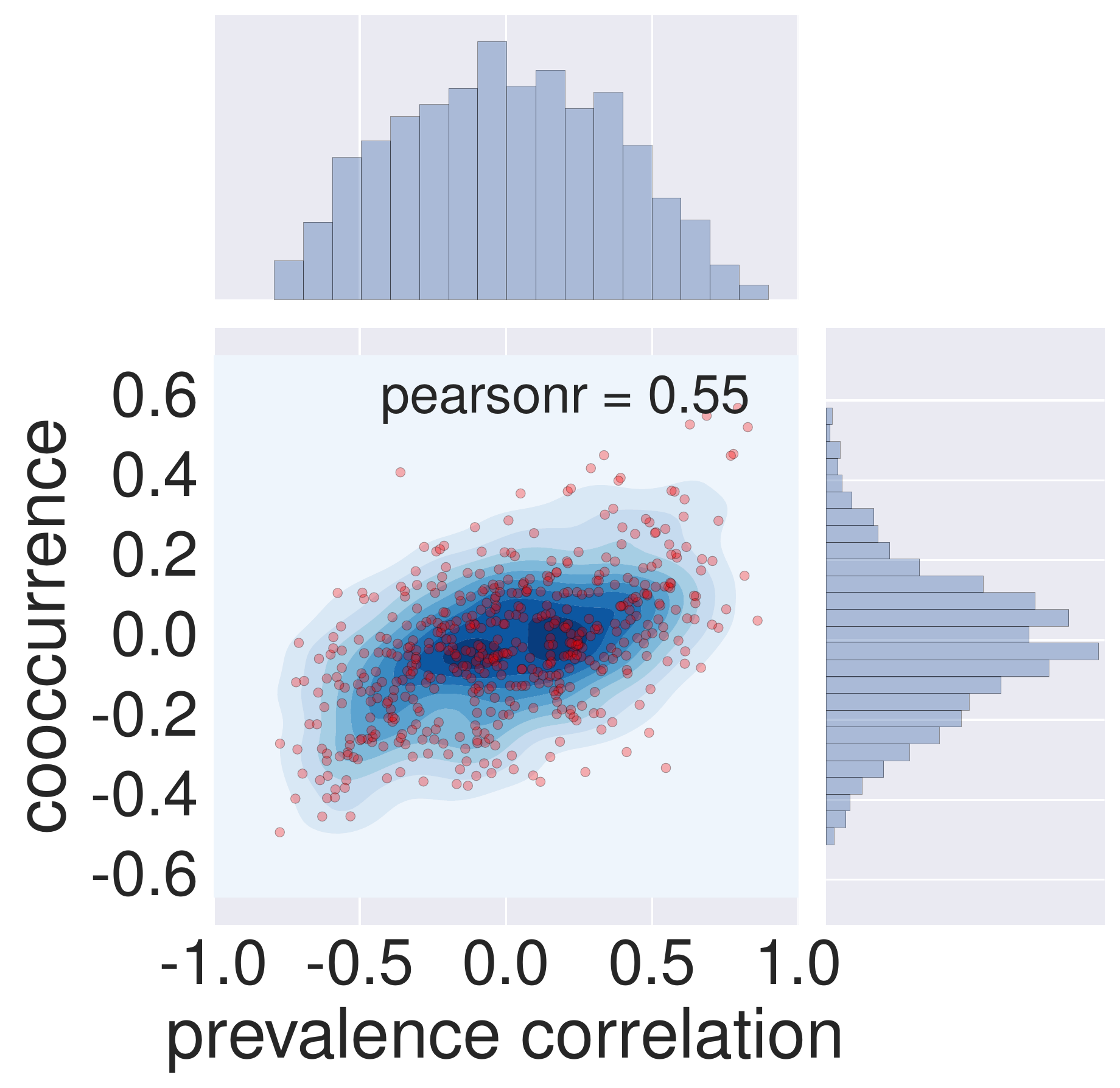}
        \caption{Immigration topics}
        \label{fig:dist_immigration_lda}
    \end{subfigure}
    \hfill
    \begin{subfigure}[t]{0.32\textwidth}
        \addFigure{\textwidth}{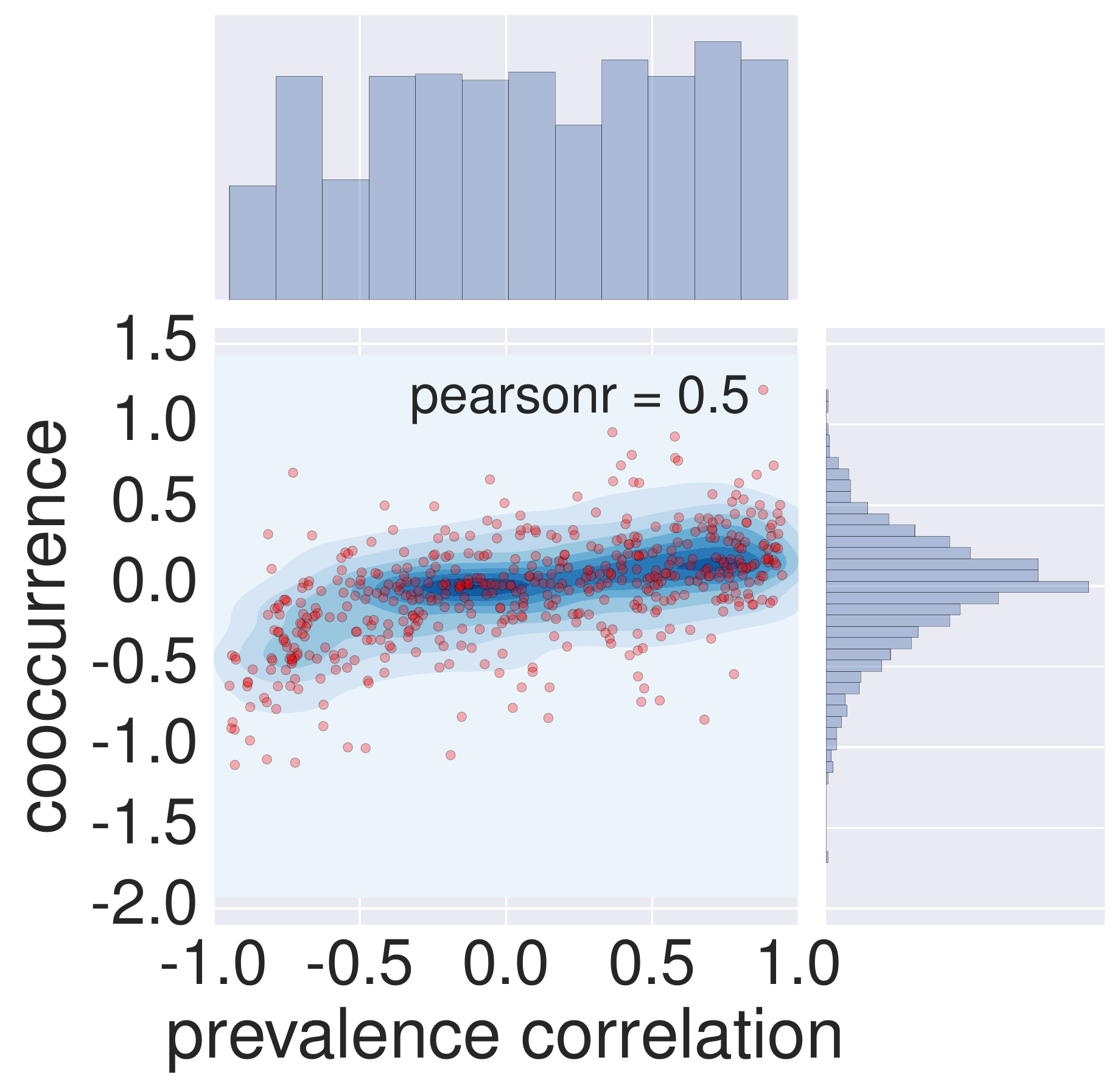}
        \caption{ACL topics}
        \label{fig:dist_acl_lda}
    \end{subfigure}
    \hfill
    \caption{Overall distributions of \cooccurrence and \correlation.
    In the main plot, each point represents a pair of \ideas:
    color density shows the kernel density estimation of the joint distribution \cite{scott2015multivariate}.
    The plots along the axes show the marginal distribution of the corresponding dimension.
    In each plot, we give the Pearson correlation, and all Pearson correlations' $p$-values are less than $10^{-40}$.
    In these plots, we use topics to represent \ideas. 
    }
    \label{fig:dist}
\end{figure*}

\subsection{Datasets and Representation of Ideas}
\label{sec:data}

We use two types of datasets to validate our framework: news articles and research papers.
We choose these two domains
because 
competition between ideas has received significant interest in history of science \cite{Kuhn:1996} and research on framing \cite{Chong:JournalOfCommunication:2007,Entman:JournalOfCommunication:1993,gitlin,lakoff2014all}.
Furthermore, 
interesting differences may exist in these two domains
as news evolves with external events and scientific research progresses through innovations.

\begin{itemize}[itemsep=0pt,topsep=1pt,leftmargin=10pt]
    \item News articles. We follow the strategy in \newcite{card2015media} to 
    obtain
    news articles from LexisNexis on five issues: abortion, immigration,  same-sex marriage, smoking, and terrorism.
    We search for relevant articles using LexisNexis subject terms in U.S.~newspapers from 1980 to 2016.
    Each of these corpora contains more than 25,000 articles.
    Please refer to the supplementary material for details.

    \item Research papers. 
    We consider full texts of papers from two communities: our own ACL community captured by
    papers from ACL, NAACL, EMNLP, and TACL from 1980 to 2014 \cite{Radev&al.09a};
    and the NIPS community from 1987 to 2016.\footnote{
    \url{http://papers.nips.cc/}. 
    }
    There are 4.8K papers from the ACL community and 6.6K papers from the NIPS community.
    The processed datasets are available at \url{https://chenhaot.com/pages/idea-relations.html}. 
\end{itemize}

\smallskip
In order to operationalize \ideas in a text corpus, we consider two ways to represent ideas.

\begin{itemize}[itemsep=0pt,topsep=1pt,leftmargin=10pt]
    \item Topics. We extract topics from each document by running LDA \citep{Blei:2003:LDA:944919.944937} on
    each
     corpus $C$.
    In all datasets, we set the number of topics to 50.\footnote{We chose 50 topics based on past experience, though this could be tuned for particular applications.
    Recall that our framework is to analyze {\em relations} between \ideas, so this choice is not essential in this work.
    }
    Formally, $\formalideas$ is the 50 topics learned from the corpus,
    and each document is represented as the set of topics 
    that are present with greater than $0.01$ probability in the topic distribution for that document.
    \item \Kwords. 
    %
    We 
    identify a list of distinguishing
      \kwords for each corpus by comparing its word frequencies to the
      background frequencies found in other corpora
      using 
      the informative Dirichlet prior model
      in \citet{Monroe21092008}.
    We set the number of \kwords to 100 for all corpora.
    For news articles, the background corpus for each issue is comprised of all articles from the other four issues.
    For research papers, we use NIPS as the background corpus for
    ACL and vice versa to identify what are the core concepts for each of these research areas. 
    Formally, $\formalideas$ is the 100 top distinguishing \kwords in the corpus and each document is represented as the set of \kwords within $\formalideas$ that 
    are present 
    in the document.
    Refer to the supplementary material for a list of example \kwords in each corpus.
\end{itemize}

In both procedures, we lemmatize all words and add common bigram
phrases to the vocabulary following
\citet{mikolov2013distributed}. %
Note that in our analysis, ideas are only present or absent in a
document, and  a document can in principle be mapped to any subset of ideas
in $\formalideas$.  In our experiments 
90\% of documents are marked
as containing
between 
7 and 
14 \ideas using topics, 8 and 33 \ideas using \kwords.

\section{Characterizing the Space of Relations}
\label{sec:observation}

To provide an overview of the four relation types in \figref{fig:intro_immigration},
we first examine the
empirical distributions of the two statistics of interest
across
pairs of ideas.
In most 
exploratory studies, however, we 
are most interested in pairs 
that exemplify each type of relation,
i.e., the most extreme points in each quadrant.
We thus look at 
these pairs in each corpus to observe how the four types 
differ in salience across datasets.

\subsection{Empirical Distribution Properties}

To the best of our knowledge, the distributions of 
pairwise
\cooccurrence and \correlation have not been examined in previous literature.
We thus first investigate the marginal distributions of \cooccurrence and \correlation and then their joint distribution.
\figref{fig:dist} shows three examples: two from news articles and 
one from research papers.
We will also focus our case studies on these three corpora in \secref{sec:explore}.
The corresponding plots for \kwords have been relegated to supplementary material due to space limitations.

\para{\Cooccurrence tends to be unimodal but not normal.}
In all of our datasets, pairwise \cooccurrence (\pmi)
presents a unimodal distribution that somewhat resembles a normal distribution, but it is rarely precisely normal. We cannot reject the hypothesis that it is unimodal for any dataset (using topics or \kwords) using the dip test \cite{hartigan1985dip}, though D'Agostino's $K^2$ test \cite{d1990suggestion} rejects normality in almost all cases. 

\para{\Correlation exhibits diverse distributions.}
Pairwise \correlation follows different distributions in news articles compared to research papers:
they are unimodal in news articles, but not in ACL or NIPS. The dip test only rejects the unimodality hypothesis in NIPS.
None follow normal distributions based on D'Agostino's $K^2$ test.

\para{\Cooccurrence is positively correlated with \correlation.}
In all of our datasets, \cooccurrence is positively correlated with
\correlation whether we use topics or \kwords to represent \ideas, although the Pearson correlation coefficients vary.
This suggests that there are 
more \friends and \headtohead relations than \romance and \armsrace relations.
Based on the results of kernel density estimation, we also observe
that this correlation is often loose, e.g., in 
ACL topics, \cooccurrence spreads widely at each mode of \correlation.

\begin{figure*}[t]
    \centering
    \begin{subfigure}[t]{0.48\textwidth}
        \addFigure{\textwidth}{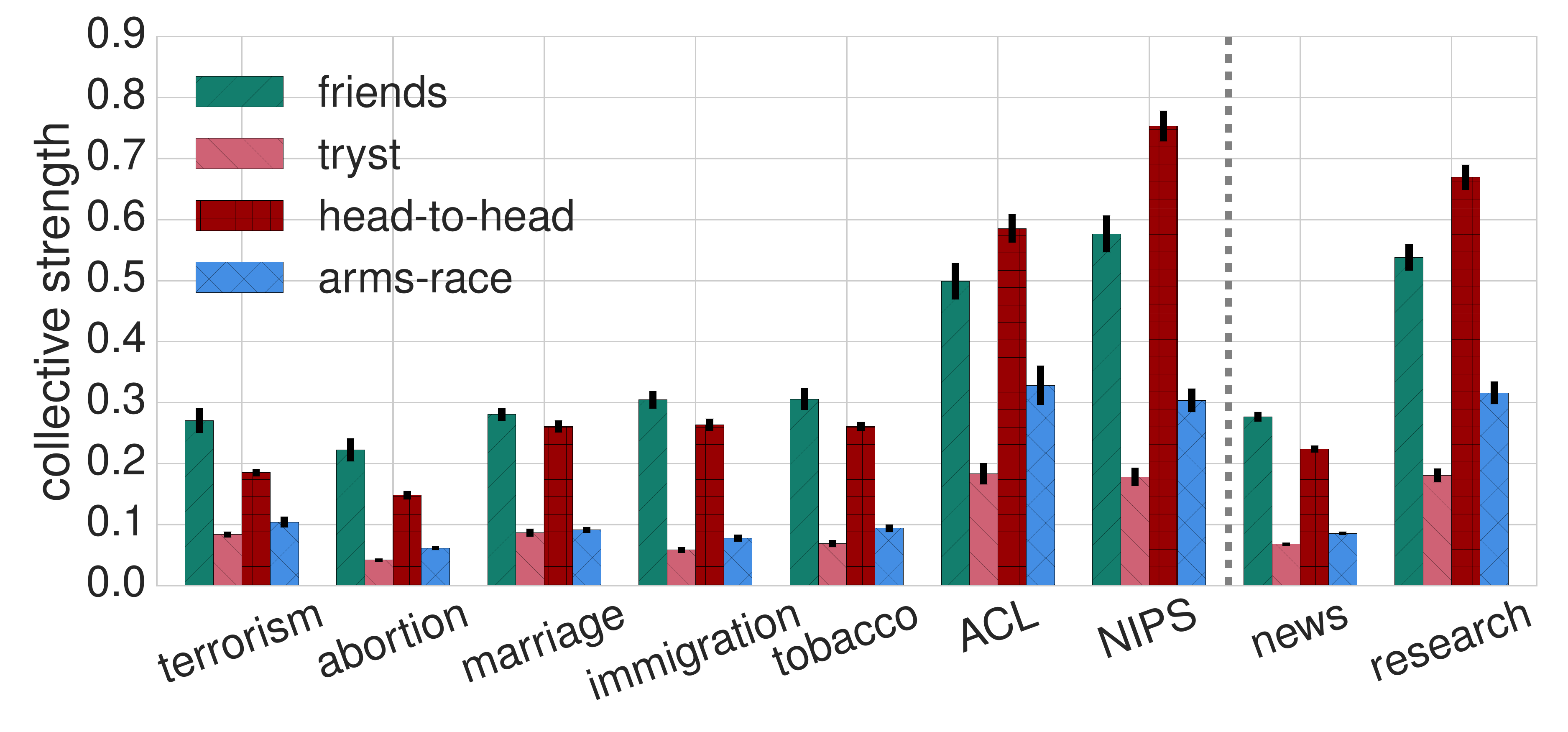}
        \caption{Topics}
        \label{fig:strength_topics}
    \end{subfigure}
    \begin{subfigure}[t]{0.48\textwidth}
        \addFigure{\textwidth}{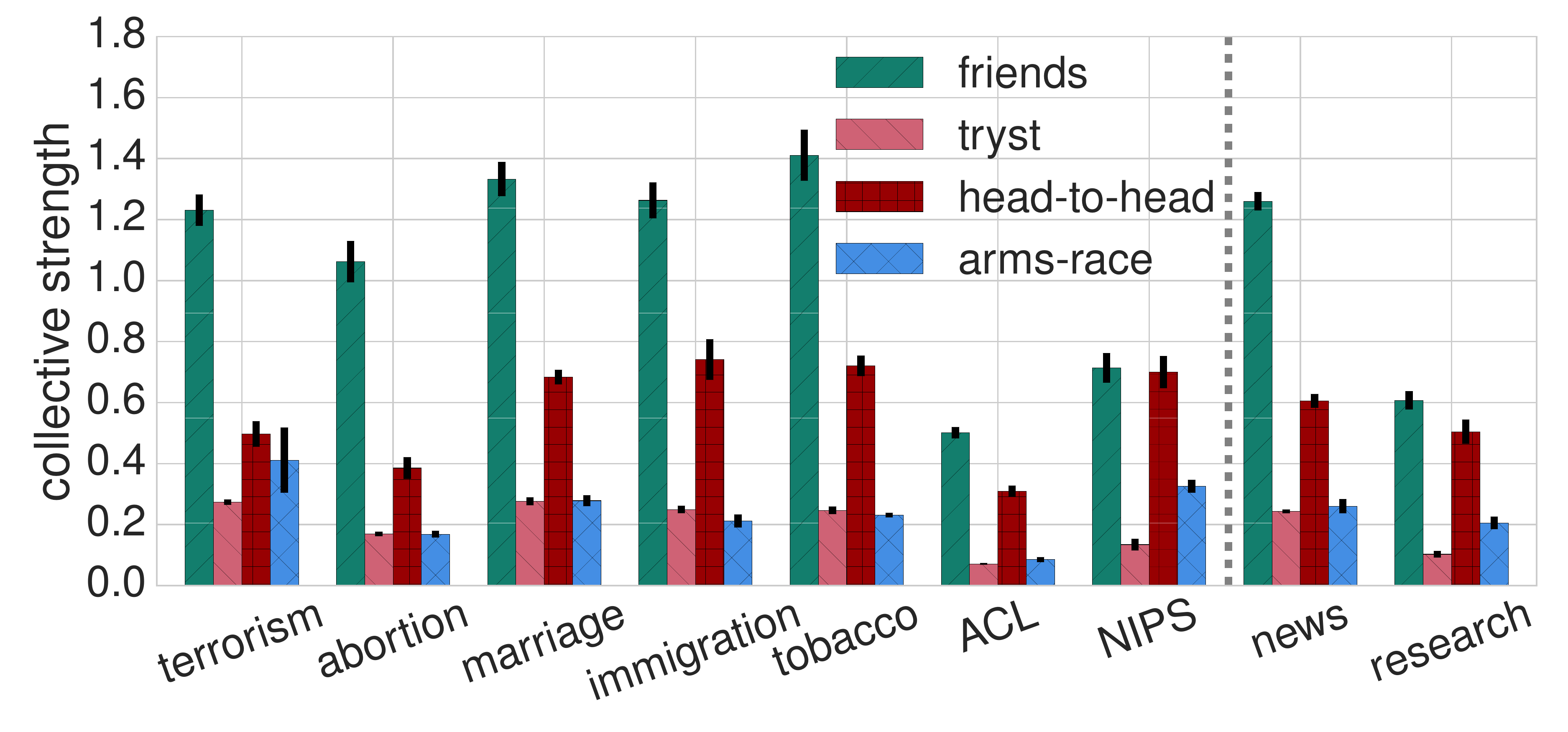}
        \caption{\Kwords}
        \label{fig:strength_words}
    \end{subfigure}
    \caption{Collective strength of the four relation types 
    in each dataset (\emph{news} is the average of the news corpora and \emph{research} is for ACL and NIPS).
    \figref{fig:strength_topics} uses topics to represent \ideas, while
    \figref{fig:strength_words} uses \kwords to represent \ideas.
    Each bar presents the average strength of the top 25 pairs in a relation type in the corresponding dataset.
    Error bars represent standard errors calculated in the usual way,
    but note that since the top 25 pairs are not random samples, they
    cannot be interpreted in the usual way.  
    %
    %
    %
    %
    }
    \label{fig:strength}
\end{figure*}

\subsection{Relative Strength of  Extreme Pairs}

%
%
%

We are interested in how our framework can identify intriguing relations between \ideas.
These potentially interesting pairs 
likely correspond to the extreme points in
each quadrant instead of the ones around the origin,
where PMI and \correlation are both close to zero. 
Here we compare the relative strength of extreme pairs in each dataset.
We will discuss how these extreme pairs 
confirm existing knowledge and suggest new hypotheses 
via case studies in \secref{sec:explore}.
For each relation type, we average the strengths of the 25 pairs with the strongest relations in that type, with
strength defined in \equationref{eq:strength}.  
This heuristic (henceforth {\em collective strength}) allows us to collectively compare the strengths of the most
prominent \friends,
\romance,
\armsrace, 
and \headtohead relations.
The results are not sensitive 
to 
the choice of 25.

\figref{fig:strength} shows the collective strength of the four 
types in all of our datasets.
The most common ordering is:
$$\text{\friends} > \text{\headtohead} > \text{\armsrace} > \text{\romance}.$$
The fact that \friends and \headtohead relations are strong is consistent with the positive correlation between \cooccurrence and \correlation.
In news, \friends is the strongest relation type, but \headtohead is the strongest in ACL topics and NIPS topics.
This suggests, unsurprisingly, that there are stronger \headtohead
competitions (i.e., one \idea takes over another) between \ideas in scientific
research than in news.  We also see that topics show
greater strength in our scientific article collections, while \kwords
dominate in news, especially in \friends.  We conjecture that terms in
scientific literature are often overloaded (e.g., a \emph{tree} could
be a parse tree or a decision tree), necessitating some abstraction
when 
representing ideas.  In contrast, news stories are more
self-contained and seek to employ consistent
usage. 

\section{Exploratory Studies}
\label{sec:explore}

We present case studies based on strongly related pairs of \ideas in the four types of relation.
Throughout this section, ``rank'' refers to the rank of the relation strength between a pair of \ideas in its corresponding relation type.

\begin{figure*}[t]
    \begin{subfigure}[t]{0.45\textwidth}
        \resizebox {\textwidth}{!} {

\begin{tikzpicture}
\node[inner sep=4pt](0) at (8, 0) {\large pakistan, india};
\node[inner sep=4pt](2) at (4, 2.4) {\large federal, state};
\node[inner sep=4pt](3) at (8, 4.8) {\large afghanistan, taliban};
\node[inner sep=4pt](4) at (0, 0) {\large israel, palestinian};
\node[inner sep=4pt](1) at (0, 4.8) {\large iran, libya};

\draw [armsracecolor, ultra thick] (0) -- (2) node[pos=.5,sloped,above] {arms-race (\#5)};
\draw [friendscolor, ultra thick] (0) -- (3) node[pos=.5,sloped,above] {friends (\#1)};
\draw [headtoheadcolor, ultra thick] (1) -- (2) node[pos=.5,sloped,above] {head-to-head (\#2)};
\draw [friendscolor, ultra thick] (1) -- (4) node[pos=.5,sloped,above] {friends (\#8)};
\draw [armsracecolor, ultra thick] (2) -- (3) node[pos=.5,sloped,above] {arms-race (\#2)};
\draw [headtoheadcolor, ultra thick] (2) -- (4) node[pos=.5,sloped,above] {head-to-head (\#11)};
\end{tikzpicture}

        }
        \caption{\raggedright Relations between a United States topic and international topics.}
        \label{fig:terrorism_relation}
    \end{subfigure}
    \hfill
    \begin{subfigure}[t]{0.26\textwidth}
        \addFigure{\textwidth}{data_exp/terrorism/{terrorism_lda_arms-race_state,federal,c_afghanistan,tal}.pdf}
        \caption{\raggedright \ideapair{``federal, state''}{``afghanistan, taliban''}}
        \label{fig:federal_taliban}
    \end{subfigure}
    \begin{subfigure}[t]{0.26\textwidth}
        \addFigure{\textwidth}{data_exp/terrorism/{terrorism_lda_head-to-head_state,federal,c_iran,american,t}.pdf}
        \caption{\raggedright \ideapair{``federal, state''}{``iran, libya''}}
        \label{fig:federal_iran}
    \end{subfigure}
    \caption{\figref{fig:terrorism_relation} shows the relations between the \ideaname{``federal, state''} topic and four international topics.
    Edge colors indicate relation types and the number in an edge label presents the ranking of its strength in the corresponding relation type.
    \figref{fig:federal_taliban} and \figref{fig:federal_iran} represent concrete examples in \figref{fig:terrorism_relation}:
    \ideaname{``federal, state''} and \ideaname{``afghanistan, taliban''} follow similar trends, although \ideaname{``afghanistan, taliban''} fluctuates over time due to significant events such as the September 11 attacks in 2001 and the death of Bin Laden in 2011; while \ideaname{``iran, lybia''} is negatively correlated with \ideaname{``federal, state''}.
    In fact, more than 70\% of terrorism news in the 80s contained the \ideaname{``iran, lybia''} topic.
    }
    \label{fig:terrorism_all}
\end{figure*}
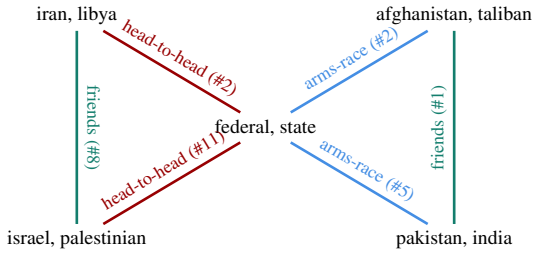
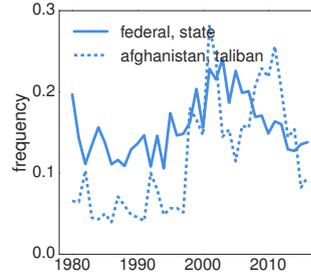
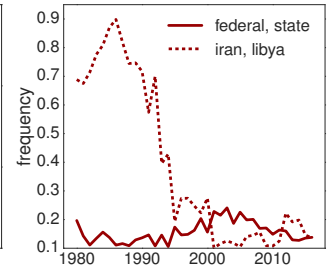
\subsection{International Relations in Terrorism}

Following a decade of declining violence in the 90s, the events of September 11, 2001 precipitated a dramatic increase in concern about terrorism, and a major shift in how it was framed \cite{kern2003}. 
As a showcase, we consider a topic which encompasses much of the
U.S.~government's
response to 
terrorism: 
\ideaname{``federal, state''}.\footnote{As in \secref{sec:intro}, we summarize each topic using a pair of strongly associated words, instead of assigning a name.}
We observe two topics engaging in an ``arms race'' with this one:
\ideaname{``afghanistan, taliban''} and \ideaname{``pakistan,
  india''}. These correspond to two geopolitical regions closely
linked to the U.S.~government's concern with terrorism, and both were
sites of U.S.~military action during the period of our dataset.
Events abroad and the U.S.~government's responses follow the \armsrace
pattern, each holding increasing attention with the other, likely because they share the same underlying cause.

We also observe two \headtohead rivals 
to the 
\ideaname{``federal, state''}
topic:  \ideaname{``iran, libya''} and \ideaname{``israel, palestinian''}.  While these topics
correspond to regions that are hotly debated in the \USA,
their coverage in news tends not to correlate temporally
with the U.S.~government's responses to terrorism,
at least during the
time period of our corpus.  Discussion of these regions was more prevalent in the
80s and 90s, with declining media coverage since 
then 
\cite{kern2003}.

The relations between these topics are consistent with structural balance theory \cite{cartwright1956structural,heider1946attitudes},
which suggests that the enemy of an enemy is a friend.
The \ideaname{``afghanistan, taliban''} topic has the strongest \friends relation with the \ideaname{``pakistan, india''} topic, i.e., they are likely to \cooccur and are positively correlated in \frequency.
Similarly, the \ideaname{``iran, libya''} topic is a close ``friend''
with the \ideaname{``israel, palestinian''} topic (ranked 8th
in \friends).

\begin{figure}[t]
    \centering
    \addFigure{0.4\textwidth}{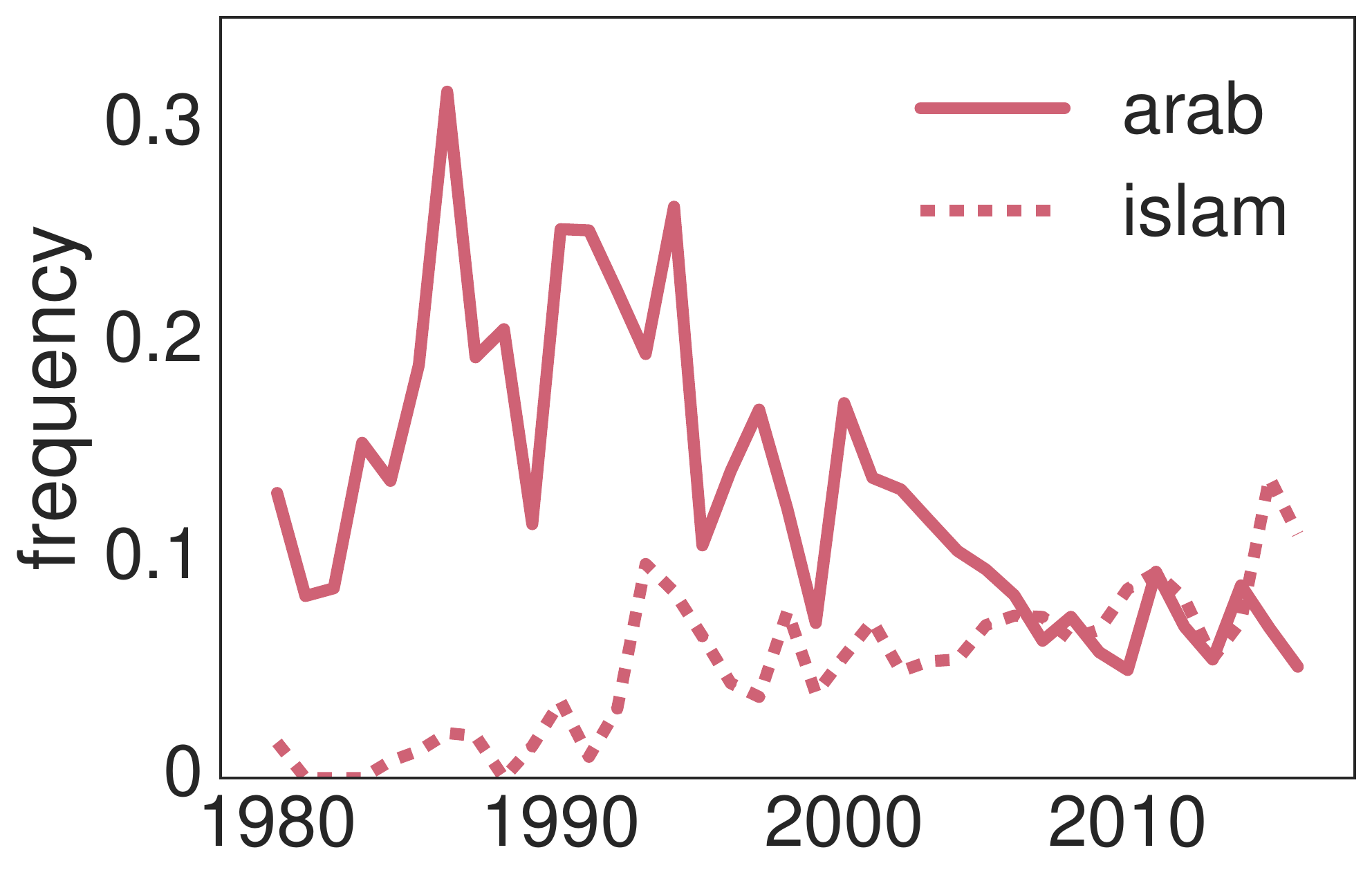}
    \caption{\Romance relation between \ideaname{arab} and
      \ideaname{islam} using \kwords to represent \ideas (\#2 in \romance%
      ): these two words tend to \cooccur but are anti-correlated in \frequency over time. In particular, \ideaname{islam} was 
      rarely used in coverage of terrorism in the 1980s.}
    \label{fig:romance_arab_islam}
\end{figure}

When using \kwords 
to represent \ideas, we observe similar relations between the term \ideaname{homeland security} and terms related to the above foreign countries.
In addition,
we highlight an interesting but unexpected \romance relation between \ideaname{arab} and \ideaname{islam} (\figref{fig:romance_arab_islam}).
It is not surprising that these two words tend to \cooccur in the same news articles, but the usage of \ideaname{islam} in the news is increasing while \ideaname{arab} is declining.
The increasing prevalence of \ideaname{islam} and decreasing prevalence of \ideaname{arab} over this time period can also be seen, for example, using Google's n-gram viewer, but it of course provides no information about cooccurrence.

This trend has not been
previously noted
to the best of our knowledge, although an article in the \emph{Huffington Post} called for news editors to distinguish Muslim from Arab.\footnote{\url{http://www.huffingtonpost.com/haroon-moghul/even-the-new-york-times-d_b_766658.html}}
Our observation suggests a conjecture that the news media have increasingly linked
terrorism to a religious group rather than an ethnic group, perhaps in
part due to the tie between the events of 9/11 and Afghanistan, which
is not an Arab or Arabic-speaking country.
We leave it to further investigation to confirm or reject this hypothesis.
%
%

%
%
%
%

\begin{table}[t]
\centering
\small

\newcolumntype{R}[1]{>{\raggedleft\arraybackslash}p{#1}}

\begin{tabular}{@{}p{0.38\textwidth}@{\hspace{6pt}}r@{\hspace{6pt}}r@{}}
\toprule
& PMI & Corr \\
\midrule
\ideaname{``federal, state''}, \ideaname{``afghanistan, taliban''} \\(\#2 in \armsrace) & 43 & 99 \\
\ideaname{``federal, state''}, \ideaname{``iran, lybia''} \\ (\#2 in \headtohead) & 36 & 56 \\
\ideaname{arab}, \ideaname{islam} (\#2 in \romance) & 106 & 1,494\\
\bottomrule
\end{tabular}

\caption{Ranks of pairs by using the absolute value of only \cooccurrence or
  \correlation.
  }
\label{tb:terrorism_pmi_corr}
\end{table}

To further demonstrate the effectiveness of our 
approach, 
we compare 
a pair's rank using only \cooccurrence or \correlation with its rank in our framework.
\tableref{tb:terrorism_pmi_corr} shows 
the results for three 
pairs above.
If we had looked at only \cooccurrence or \correlation,
we would probably have missed these interesting pairs.

\subsection{Ethnicity \Kwords in Immigration}
In addition to results on topics in \secref{sec:intro}, 
we observe unexpected patterns about 
ethnicity \kwords in immigration news.
Our observation 
starts with a top \romance relation between  \ideaname{latino} and \ideaname{asian}.
Although these words are likely to \cooccur, their \frequency trajectories differ, with the discussion of Asian immigrants in the 1990s giving way to a focus on the word \ideaname{latino} from 2000 onward. 
Possible theories to explain this observation include that undocumented immigrants are generally perceived as a Latino issue, or that Latino voters are increasingly influential in U.S. elections.

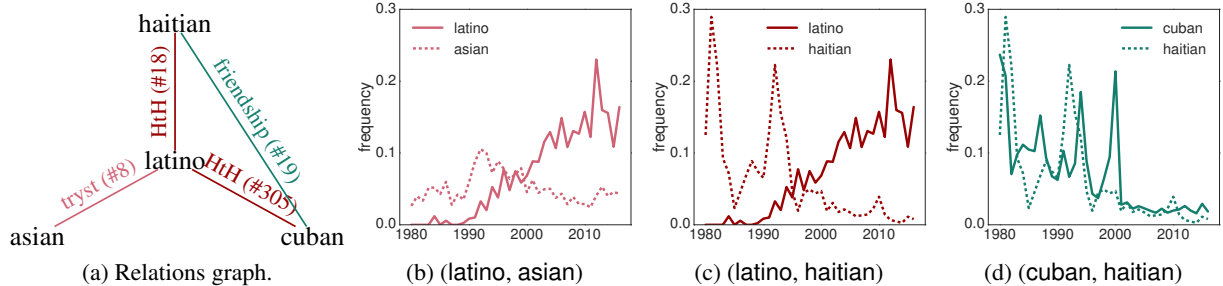
\begin{figure*}[t]
   \begin{subfigure}[t]{0.28\textwidth}
      \resizebox {\textwidth}{!} {

\begin{tikzpicture}
\node[inner sep=0pt](0) at (4.5, 2.5) {\Huge latino};
\node[inner sep=0pt](1) at (0, 0) {\Huge asian};
\node[inner sep=0pt](2) at (9, 0) {\Huge cuban};
\node[inner sep=0pt](3) at (4.5, 7) {\Huge haitian};

\draw [romancecolor, ultra thick] (0) -- (1) node[pos=.5,sloped,above] {\huge \romance (\#8)};
\draw [headtoheadcolor, ultra thick] (0) -- (2) node[pos=.5,sloped,above] {\huge HtH (\#305)};
\draw [headtoheadcolor, ultra thick] (0) -- (3) node[pos=.5,sloped,above] {\huge HtH (\#18)};
\draw [friendscolor, ultra thick] (2) -- (3) node[pos=.5,sloped,above] {\huge \friends (\#19)};
\end{tikzpicture}

      }
      \caption{%
      Relations graph.
      }
      \label{fig:imm_relation}
   \end{subfigure}
   \begin{subfigure}[t]{0.23\textwidth}
        \addFigure{\textwidth}{data_exp/immigration/{immigration_lexicon_romance_latino_asian}.pdf}
        \caption{\ideapair{latino}{asian}
        }
      \label{fig:imm_latino_asian}
   \end{subfigure}  
   \hfill
   \begin{subfigure}[t]{0.23\textwidth}
        \addFigure{\textwidth}{data_exp/immigration/{immigration_lexicon_head-to-head_latino_haitian}.pdf}
        \caption{\ideapair{latino}{haitian}
        }
      \label{fig:imm_latino_cuban}
   \end{subfigure}
   \hfill
   \begin{subfigure}[t]{0.23\textwidth}
      \addFigure{\textwidth}{data_exp/immigration/{immigration_lexicon_friends_cuban_haitian}.pdf}
        \caption{\ideapair{cuban}{haitian}
        }
      \label{fig:imm_cuban_haitian}
   \end{subfigure}
   \caption{Relations between ethnicity \kwords in immigration news (HtH for \headtohead): \ideaname{latino} holds a \romance relation with \ideaname{asian} and \headtohead relations with two subgroups from Latin America, \ideaname{haitian} and \ideaname{cuban}.
   We do not show the relations between \ideaname{asian} and \ideaname{haitian, cuban}, because their strength is close to 0.}
    \label{fig:imm_origin}
\end{figure*}

Furthermore,
\ideaname{latino} holds \headtohead relations with two subgroups of 
Latin
American
 immigrants: \ideaname{haitian} and \ideaname{cuban}.
In particular, the strength of the relation with \ideaname{haitian} is ranked \#18 in \headtohead relations.
Meanwhile, \ideaname{haitian} and \ideaname{cuban} have a \friends relation, which is again consistent with structural balance theory.
The decreasing prevalence of \ideaname{haitian} and \ideaname{cuban} 
perhaps speaks to the shifting geographical focus of recent immigration to the U.S., and issues of the Latino pan-ethnicity.
In fact, a majority of Latinos prefer to identify with their national
origin relative to the pan-ethnic terms
\cite{taylor2012labels}. However, we should also note that much of
this coverage relates to a set of specific refugee crises, temporarily
elevating the political importance of these nations in the U.S. 
Nevertheless,
the underlying social and political reasons behind these \headtohead relations
are worth further investigation.

\subsection{Relations between Topics in ACL}

\begin{figure}[t]
    \resizebox {0.48\textwidth}{!} {

\begin{tikzpicture}
\node[inner sep=3pt](0) at (4, 2.4) {\large machine translation};
\node[inner sep=3pt](1) at (0, 0) {\large rule,forest methods};
\node[inner sep=3pt](2) at (8, 4.8) {\large word alignment};
\node[inner sep=3pt](3) at (8, 0) {\large sentiment analysis};
\node[inner sep=3pt](4) at (0, 4.8) {\large discourse (coherence)};

\draw [romancecolor, ultra thick] (0) -- (1) node[pos=.5,sloped,above] {\romance (\#5)};
\draw [friendscolor, ultra thick] (0) -- (2) node[pos=.5,sloped,above] {\friends (\#1)};
\draw [armsracecolor, ultra thick] (0) -- (3) node[pos=.5,sloped,above] {\armsrace (\#1)};
\draw [headtoheadcolor, ultra thick] (0) -- (4) node[pos=.5,sloped,above] {\headtohead (\#1)};
\draw [headtoheadcolor, ultra thick] (1) -- (3) node[pos=.5,sloped,above] {head-to-head (\#38)};
\draw [armsracecolor, ultra thick] (1) -- (4) node[pos=.5,sloped,above] {\armsrace (\#23)};
\draw [armsracecolor, ultra thick] (2) -- (3) node[pos=.5,sloped,above] {\armsrace (\#2)};
\draw [headtoheadcolor, ultra thick] (2) -- (4) node[pos=.5,sloped,above] {\headtohead (\#7)};
\end{tikzpicture}

    }
    \caption{Top relations between the topics in ACL Anthology.
    The top 10 words for the \ideaname{rule, forest methods} topic are \emph{rule, grammar, derivation, span, algorithm, forest, parsing, figure, set, string}.}
    \label{fig:acl_relation}
\end{figure}
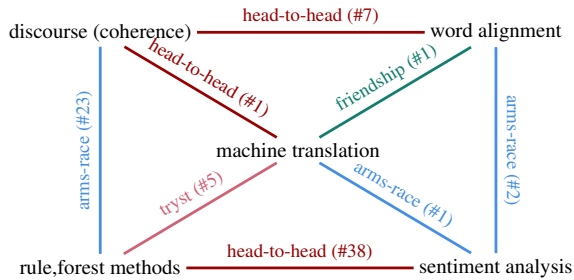

Finally, we analyze relations between topics in the ACL Anthology. 
It turns out that \ideaname{``machine translation''} is at a central position among top ranked relations in all the four types (\figref{fig:acl_relation}).%
\footnote{In the ranking, we filtered a topic where the top words are \emph{ion, ing, system, process, language, one, input, natural language, processing, grammar}.  For the purposes of this corpus, this is effectively a stopword topic.}
It is part of the strongest relation in all four types except \romance (ranked \#5).

The full relation graph presents further patterns.
\Friends 
demonstrates transitivity: both \ideaname{``machine translation''} and \ideaname{``word alignment''} have similar relations with other topics.
But such transitivity does not hold for \romance:
although the \frequency of \ideaname{``rule, forest methods''} is 
anti-correlated with both \ideaname{``machine translation''} and \ideaname{``sentiment analysis''}, \ideaname{``sentiment analysis''} seldom \cooccurs with \ideaname{``rule, forest methods''} because \ideaname{``sentiment analysis''} is seldom built on parsing algorithms.
Similarly, \ideaname{``rule, forest methods''} and \ideaname{``discourse (coherence)''} 
hold an \armsrace relation: they 
do not tend to \cooccur and both decline in 
relative \frequency as \ideaname{``machine translation''} rises.

The prevalence of each of these ideas in comparison to \ideaname{machine translation} is shown in in \figref{fig:acl}, which reveals additional detail.

\begin{figure*}[htb!]
    \centering

\begin{tabular}{c@{\hspace{0.02\textwidth}}c@{\hspace{0.02\textwidth}}|@{\hspace{0.02\textwidth}}c}
& {\Large anti-correlated} & {\large correlated} \\
\parbox[t]{2mm}{\multirow{2}{*}{\rotatebox[origin=c]{90}{\large likely to \cooccur}}} & {\large \textcolor{romancecolor}{\romance}} & {\large \textcolor{friendscolor}{\friends}} \\
& \addFigure{0.3\textwidth}{data_exp/acl/{acl_lda_romance_translation,sys_rule,grammar,de}.pdf}
& \addFigure{0.3\textwidth}{data_exp/acl/{acl_lda_friends_translation,sys_alignment,word,}.pdf}
\\
\midrule
\parbox[t]{2mm}{\multirow{2}{*}{\rotatebox[origin=c]{90}{\large unlikely to \cooccur}}} & {\large \textcolor{headtoheadcolor}{\headtohead}} & {\large \textcolor{armsracecolor}{\armsrace}}\\
& \addFigure{0.3\textwidth}{data_exp/acl/{acl_lda_head-to-head_translation,sys_discourse,relat}.pdf}
& \addFigure{0.3\textwidth}{data_exp/acl/{acl_lda_arms-race_translation,sys_sentiment,opini}.pdf}\\
\end{tabular}
    \caption{
    Relations between topics in ACL Anthology in the space of \cooccurrence and \correlation (\correlation is shown explicitly and \cooccurrence is encoded in row captions), color coded to match the text.
    The $y$-axis represents the
    relative proportion    
     of papers in a year that contain the corresponding topic.
    The top 10 words for the \ideaname{rule, forest methods} topic are
    \emph{rule, grammar, derivation, span, algorithm, forest, parsing,
      figure, set, string}. 
    \label{fig:acl}}
\end{figure*}

\section{Related Work}
\label{sec:related}

We present two strands of related studies in addition to what we have discussed.

\para{Trends in \ideas.} 
Most studies have so far examined the trends of ideas individually \cite{,Michel:Science:2011,Hall:2008:SHI:1613715.1613763,Rule:ProceedingsOfTheNationalAcademyOfSciences:2015}. 
For instance, \citet{Hall:2008:SHI:1613715.1613763} present various trends in our own computational linguistics community, including the rise of statistical machine translation.
More recently, rhetorical framing has been used to predict these sorts of patterns \cite{prabhakaran1predicting}.
An exception is that \citet{shi2010leads} use  \correlation to analyze lag relations between topics in publications and research grants.
Anecdotally, \citet{grudin2009ai} observes a ``\headtohead'' relation between artificial intelligence and human-computer interaction in research funding.
However, to our knowledge, our work is the first study to systematically characterize 
relations \emph{between} \ideas.

\para{Representation of \ideas.}
In addition to topics and \kwords, studies have also sought to operationalize the ``memes'' metaphor 
using quotes 
and text reuse in the media \cite{Leskovec2009Memes,Niculae2015QUOTUS,Smith2013Infectious,Wei2013Competing}. 
In topic modeling literature, \citet{blei2006correlated} also point out that topics do not \cooccur independently and explicitly model the \cooccurrence within documents.

\section{Concluding Discussion}
\label{sec:conclusion}

We proposed a method to characterize relations between \ideas 
in texts
through the lens of \cooccurrence within documents and \correlation over time.
For the first time, we observe that the distribution of pairwise \cooccurrence is unimodal, while the distribution of pairwise \correlation is not always unimodal,
and show that they are positively 
correlated.
This combination suggests four types of relations between ideas, and these four types are all found to varying extents in our experiments.

We illustrate our computational method by exploratory 
studies on news corpora and scientific research papers.
We not only confirm existing knowledge but also suggest hypotheses around the usage of \ideaname{arab} and \ideaname{islam} in terrorism and \ideaname{latino} and \ideaname{asian} in immigration.

It is important to note that the relations found using our approach depend on the nature of the 
representation of \ideas and the source of texts.
For instance, we cannot expect relations found in news articles to reflect shifts in public opinion if news articles do not effectively track public opinion. 

Our method is entirely observational.
It remains as a further stage of analysis to understand the underlying reasons that lead to these relations between \ideas.
In scientific research, for example, it could simply be the progress of science,
i.e., newer \ideas overtake older ones deemed less valuable at a given
time;
on the other hand,
history suggests that it is not always the correct \ideas that
are most expressed, and many other factors may be important.
Similarly, in news coverage, 
underlying sociological and political situations have significant impact on which \ideas are presented, and how.

There are many potential directions to improve our 
method to account for complex relations between ideas.
For instance, we assume that both \ideas and relations are statically
grounded in \kwords or topics.
In reality, ideas and relations both evolve over time:
a \romance relation might appear as \friends if we focus on a
narrower time period.
Similarly, new \ideas show up and even the same \idea may change over time and be represented by different words.
    \para{Acknowledgments.}
    We thank Amber Boydstun, Justin Gross, Lillian Lee, anonymous reviewers, and all members of Noah's ARK for helpful comments and discussions.
This research was made possible by a Natural Sciences
and Engineering Research Council of Canada Postgraduate
Scholarship (to D.C.) and a University
of Washington Innovation Award.

	\bibliography{ref}
	\bibliographystyle{acl_natbib}

\end{document}